\documentclass[twocolumn,preprintnumbers,amsmath,amssymb]{revtex4}

\usepackage{amsmath}
\usepackage{amsfonts}
\usepackage{amssymb}
\usepackage{graphicx}

\newcommand{\be}{\begin{equation}}
\newcommand{\ee}{\end{equation}}
\newcommand{\bea}{\begin{eqnarray}}
\newcommand{\eea}{\end{eqnarray}}
\newcommand{\OO}{\Omega_{\phi 0}}

\begin{document}
\title{\large \bf Measuring deviations from a cosmological constant: \\ 
a field-space parameterization}

\author{\normalsize Robert Crittenden$^1$, Elisabetta Majerotto$^1$ and Federico Piazza$^{1,2}$ \\
\normalsize \it $^1$Institute of Cosmology and Gravitation, University of Portsmouth, UK\\
\normalsize \it $^2$ Perimeter Institute for Theoretical Physics, Waterloo, Ontario}


\begin{abstract}
Most parameterizations of the dark energy equation of state do not reflect
realistic underlying physical models.  Here, we develop a relatively simple 
description of dark energy based on the dynamics of a scalar field which is 
exact in the limit that the equation of state approaches a cosmological constant, 
assuming some degree of smoothness of the potential.  
By introducing just two parameters defined in the configuration space of the field
we are able to reproduce a wide class of quintessence models.
We examine the observational constraints on these models as compared to linear 
evolution models, and show how priors in the field space translate into priors on 
observational parameters. 
\end{abstract}

\maketitle

\setcounter{footnote}{0} \setcounter{page}{1}
\setcounter{section}{0} \setcounter{subsection}{0}
\setcounter{subsubsection}{0}


Studies of dark energy \cite{shin} are hindered by the lack of specific alternatives to the cosmological constant model.  Once we allow that the dark energy density might vary in time, how it evolves (most often parameterised by the equation of state $w(z)$) appears to be virtually a free function in most scenarios, making the models difficult to constrain.
As a result, the dynamical dark energy evolution is often quantified instead by somewhat arbitrary phenomenological models, such as assuming that $w(z)$ evolves linearly in redshift or the 
scale factor.  However, such models generally have little motivation from fundamental physics, and can often act in ways (such as crossing the $w=-1$ `phantom divide,' thereby violating the weak energy condition) which can be difficult to explain 
from an underlying physics model.

One commonly invoked physical model for dark energy is quintessence, where a scalar field $\phi$ rolling down its potential comes to dominate the energy density of the Universe \cite{Wetterich,Ratra}.  In such models, the evolution of the equation of state is determined by the potential $V(\phi)$. It has been shown that any evolution of 
the equation of state could result from a suitably chosen potential \cite{Steinhardt}; however, it does not necessarily follow that all possible dark energy histories are equally likely.  Here we examine the question of 
what priors on the functional space of $w(z)$ result from the scalar field model, and develop a two parameter description which captures the behaviour of quintessence when its energy density is slowly evolving.

If we take dark energy to be effectively described by a scalar field $\phi$, it seems reasonable to 
expect some level of smoothness in the configuration space of the field. 
For instance, we might expect that relevant variations of the Lagrangian to occur only over field-space distances of order one in some ``smoothness scale'' $M$. In the absence of univocal theoretical indications, there appears no strong motivation for choosing any particular scale for $M$; however, the observational requirements of quintessence naturally associate it with the Planck mass. 
The general requirements for the potential $V$ of a minimally coupled 
quintessence field are that it be of the size of the present energy density, 
$V(\phi_0) \sim M_P^2 H_0^2$ (where $M_P = (8 \pi G)^{-1/2}$ is the reduced Planck mass) and its second 
derivative to be as small as the Hubble scale, $V''(\phi_0) \lesssim H_0^2$, so that the field is light enough to still be evolving today. We can introduce the smoothness scale $M$ by writing the potential as $V(\phi) = M_P^2 H_0^2 f(\phi/M)$ and by assuming the 
``well behaved'' function $f$ and its derivatives to be at most of order one. 
The condition  $V''(\phi_0) \lesssim H_0^2$ is then naturally satisfied if $M \gtrsim M_P$.

The observational status of the dark energy 
equation of state provides the other basic motivation for this work. Recently, $w(z)$
has been confined \cite{Astier:2005qq,Spergel,Riess,Wood} to a region of the 
parameter space relatively close to the cosmological constant behaviour $w = -1$. 
For a minimally coupled scalar field this corresponds to a slow-roll regime
$\dot\phi^2/2 \ll V$ which, in turn, implies that the region  $\Delta \phi$ probed by the field at 
late times is in fact small in Planck units. 

For these reasons, it is interesting to find a field-space parameterization of dark energy. 
In slow-roll inflation it is customary to assume that the dynamics is independent 
of the initial conditions and one defines field-space quantities (the slow-roll parameters) which 
capture the substantial features of the cosmic evolution. The presence of a significant dark matter  component makes things more difficult in the case of quintessence. Working in units where $M_P^2 = 2$, the basic equations are
\bea
\ddot  \phi +  3 H \dot \phi + V'(\phi) = 0,  \label{1}\\  
6 H^2 \,  =\, \rho_m + \rho_\phi, \label{2}
\eea 
 where $\rho_\phi = \dot \phi^2/2 + V(\phi)$ and $\rho_m$ are the scalar and 
non-relativistic matter energy densities respectively. The size of the acceleration
of the field relative to the damping term, 
$
\beta \equiv {\ddot  \phi}/{3 H \dot \phi},
$ 
is no longer systematically negligible. Typically, in thawing models discussed below, while the universe remains matter  dominated $\beta = 1/2$, which then begins to decline towards zero when the quintessence field begins to dominate (see also \cite{linder2} about this). By Eq. \eqref{1}, the deviation from $w = -1$ is given by  
\be
1 + w = \frac {\dot \phi ^ 2 }{ \rho_\phi} = \frac{2}{3}\ \frac{V'^2}{6H^2(1+\beta)^2 \rho_\phi}.
\ee

Instead of using the usual slow-roll parameter $\epsilon \equiv (V'/V)^2$, we can try to reproduce the quintessence dynamics by formally considering $\beta$ as a function of $\phi$ and 
making an approximation for the function 
\be
\kappa(\phi) \, \equiv \, \frac{V'}{V (1+\beta)} .
\ee
The following equations are easily found:
\bea
 1 + w & = & \kappa^2(\phi) \Omega_\phi (1 - w)^2 /6 \label{exact1}\\
 \frac{d\phi}{d \ln a} & = & - \kappa(\phi) \Omega_\phi (1 - w),   \label{exact2}
\eea  
where $\Omega_\phi = \rho_\phi /(\rho_m + \rho_\phi) $ and $a = 1/(1+z)$ is the scale factor. 
The above equations are as yet exact. 
When the dark energy density is nearly constant in time,  $1 + w \ll 1$ and Eq. (\ref{exact1}) implies that $\kappa(\phi$) is also small.  We can then consistently drop terms of order ${\cal{O}}(1+w)$ and we can substitute $\Omega_\phi$  
with its ``zeroth order'' approximation i.e. the fractional energy density corresponding to a cosmological constant,
\begin{equation}
 \label{5}
\Omega_\Lambda (a) = \frac{\OO} {\OO + (1 - \OO) a^{-3}},
\end{equation}
$\OO$ being the fractional energy density of dark energy today.  
(The differences in between $\Omega_\Lambda (a)$ and the true fractional density 
$\Omega_\phi(a)$ are actually of order 
$\int^1_a (1+w) da'/a' $, but this should be no larger than ${\cal{O}}(1+w)$ for the redshift range where the dark energy is dynamically interesting.)  
Equations \eqref{exact1} and \eqref{exact2} can then be written as 
\bea
 1 + w & = & {2}\kappa^2(\phi)  \Omega_\Lambda(a) /3 \label{w}\\
 \frac{d\phi}{d \ln a} & = & - 2 \kappa(\phi) \Omega_\Lambda(a) .   \label{n-vel}
\eea 

The above equations, combined with an expansion for $\kappa(\phi)$, define our parameterization.
There is still a one to one correspondence between the functions $\kappa(\phi)$ and $w(a)$;
given a particular $\kappa(\phi)$ one can solve for $w(a)$ and vice versa.  However, the requirements 
of smoothness of the potential constrain the evolutions that are produced.  While the full potential
still has an infinite number of parameters, the small field displacement when $\kappa(\phi)$ is small 
ensures that only a narrow region of the field space is actually probed by observations. 
The potential elsewhere is irrelevant;
its associated parameters do not affect the likelihood and can be trivially integrated 
over in the evidence calculation, yielding an effective theory which has very few parameters. 
Note that the number of effective parameters of the model will depend on the values of the parameters themselves. For example, the possibility that the field sits at rest in a local minimum corresponds here to $\kappa(\phi_0) = 0$. In this case, the model trivially reduces to a cosmological constant regardless of the values of $\kappa$ at $\phi \neq \phi_0$.

We found that a linear approximation for  
$\label{d(phi)}
\kappa(\phi) = \kappa_0 + \kappa_1 (\phi - \phi_0)$
already captures very well the dynamics of models reasonably close to a cosmological 
constant. 
By inserting this linear form into \eqref{n-vel} and integrating, we find $\phi - \phi_0$
as a function of $a$. This gives
\begin{equation} \label{evo2}
\kappa(a) = \kappa_0 (1 - \OO + \OO a^{3})^{-2 \kappa_1/3}
= \kappa_0 \left[\frac{\Omega_\Lambda(a)}{a^3 \OO} \right]^{2 \kappa_1/3}.
\end{equation}
This and Eq. \eqref{w} specify uniquely $w(a)$, which then can be integrated for the resulting evolution of the dark energy 
density $\rho_{\phi} \propto \exp[\mathcal{I}(a)]$, where 
 \bea \mathcal{I}(a) &=& 3\int^1_a \frac{da'}{a'}\, ({1+w(a')})
 \simeq 2\int^1_a \frac{da'}{a'} \, {\kappa^2(a')} \Omega_{\Lambda}(a')\nonumber \\
  &=&(\kappa^2(a) - \kappa_0^2)/{2\kappa_1}. \label{evo3}
 \eea
Thus the dark energy density takes a very simple evolution in terms of 
$ \Omega_{\Lambda}(a)$. It is straightforward to check that our model is still well defined 
when $\kappa_1 \rightarrow 0$ and $\kappa(a)$ is effectively constant. 

Quintessence models have been separated into two basic classes \cite{linder}: those in which the field is effectively fixed at early times and only begins to evolve recently (``thawing'' models) and those in which the field rolls quickly down a potential at early times, and only now is starting to slow down as the potential flattens out (``freezing'' models.)  In the former, the equation of state begins at $w=-1$ and rises at late times.  In the latter, the equation of state begins at some higher value and drops towards $w=-1$ at late times.  It is interesting to see how well our parameterization works for these models.   
\begin{figure}[htbp] 
\begin{center} \vspace{-1.0cm}
\includegraphics*[width=3.2in]{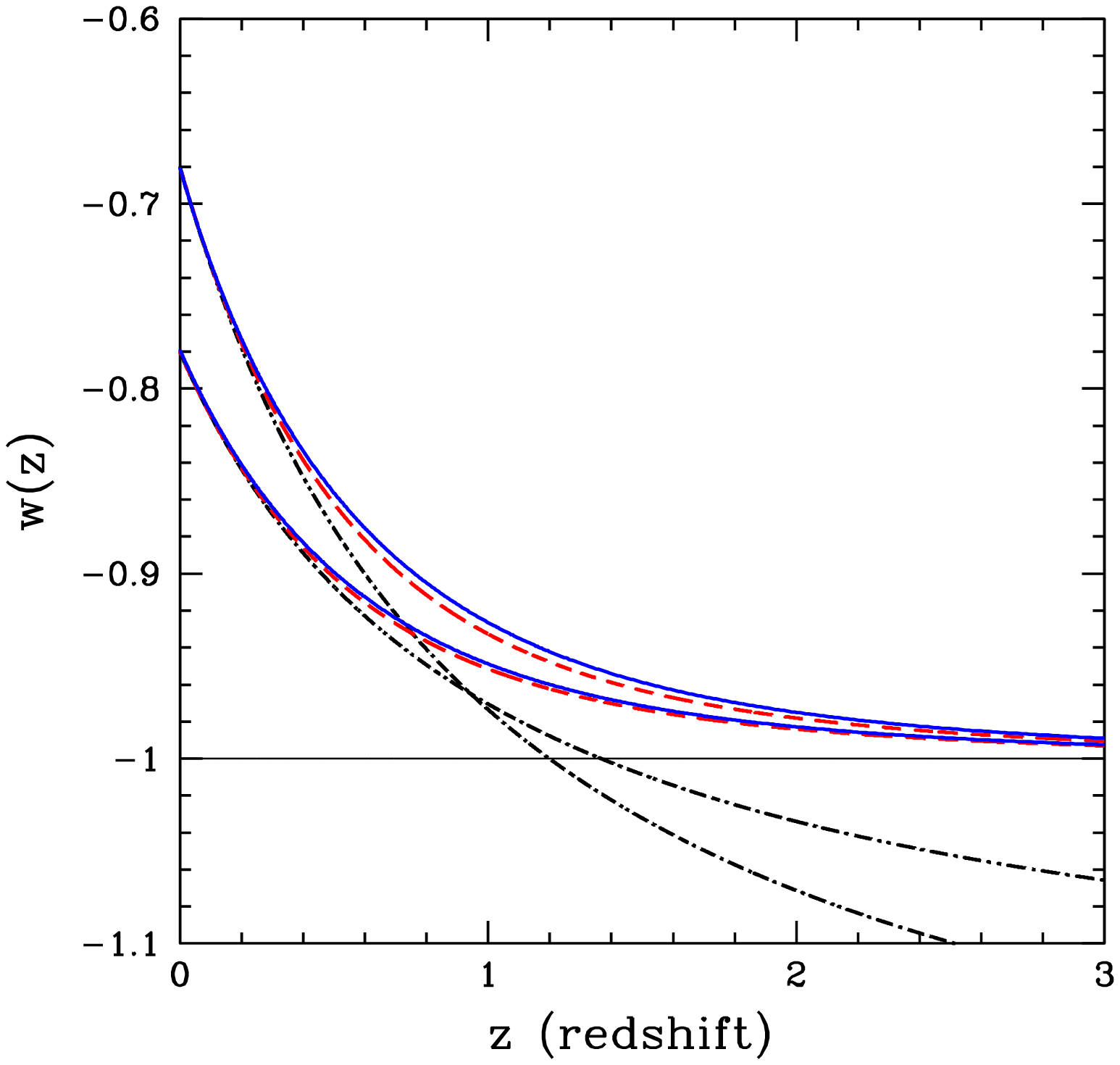}%
\hfil
\vspace{-1.5cm}
\includegraphics*[width=3.2in]{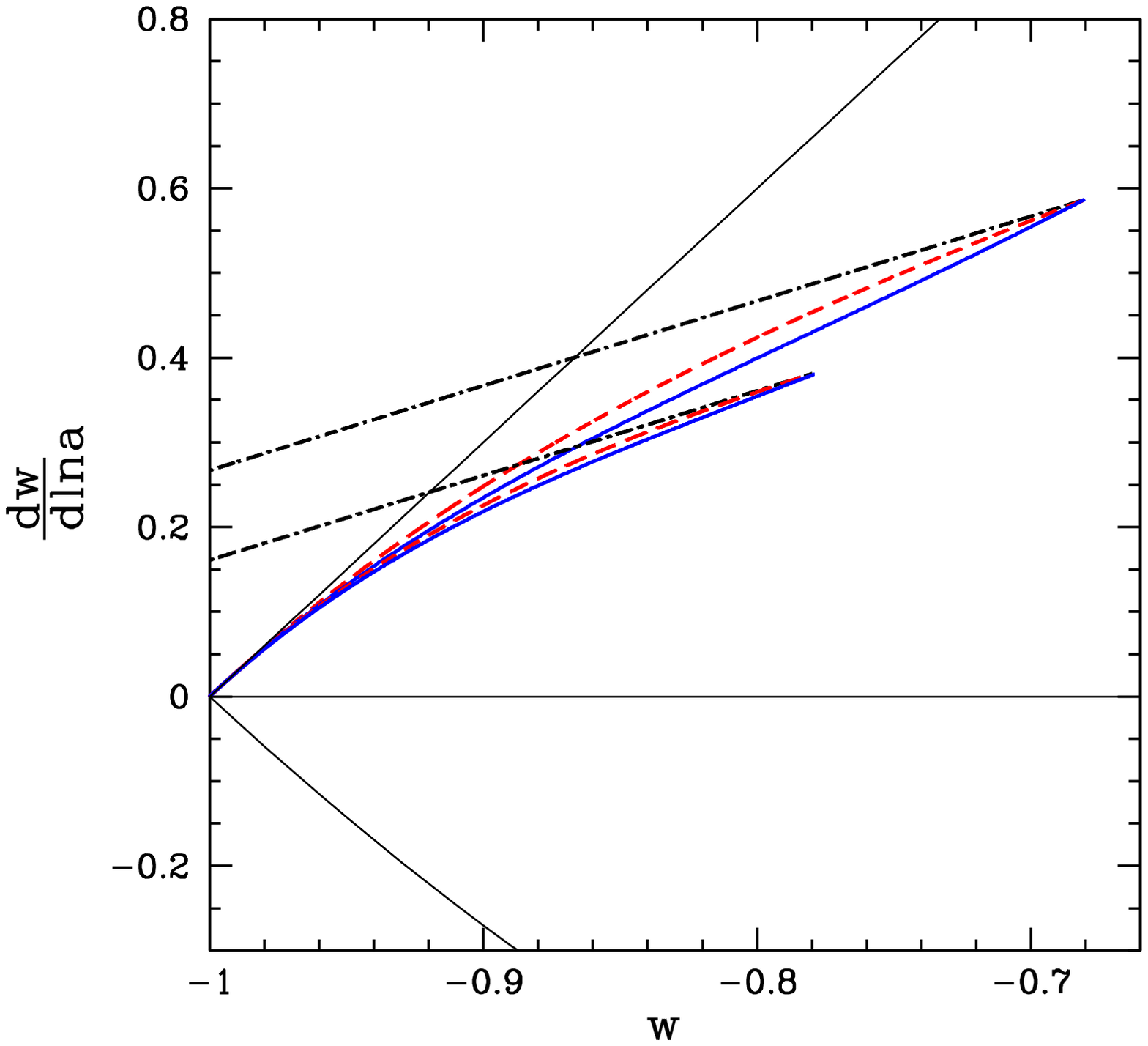}%
\end{center}\vspace{-1cm}
\caption{
The evolution of dark energy equation of state as a function of redshift, and in the 
$(w,w')$ plane for different parameterizations.  The solid (blue) lines show the results 
of an exact integration for a typical potential (quadratic in this case), while the dashed (red) lines show our parameterization.  The approximation improves as the present equation of state approaches $-1$.  Also shown (dot-dashed) are similar curves for a linear parameterization, $w(a) = w_0 + w_a(1-a)$.  We chose parameters to match $w$ and $w'$ at the present.     
}\vspace{-.4cm}
\label{fig:thaw} \end{figure}

In Figure \ref{fig:thaw} we plot the cosmological evolution of the dark energy equation
of state for different values of $\kappa_0$ and $\kappa_1$.
Our ``smoothness requirement'' is generally met for the potentials of the thawing models, and therefore our parameterization does very well at reproducing their behaviour.  
We find that the two parameters cover very well the entire range of the ``thawing'' quintessence models.
On the other hand, freezing models, while interestingly addressing the 
``coincidence problem'' \cite{coin}, typically exhibit singular potentials and rather 
fine-tuned conditions for the field.
Steep curvature and rapidly changing potentials at early times are in fact required
in order to maintain a tracking \cite{track} behaviour, and so are not well described 
by our parameterization. 
 
Our approach results in a two parameter $(\kappa_0, \kappa_1)$ family of $w(a)$ histories which can be used to fit the data, much like the often-used linear parametrization, $w(a) = w_0 + w_a(1-a)$ \cite{Chevallier,Linder:2002}.  If we use our parametrization to fit a fully dynamical thawing model, the values of $(\kappa_0, \kappa_1)$ we obtain should reflect the true $\kappa(\phi)$, modulo corrections of order $1+w$. 

To simplify comparison to the data and past literature \citep{linder,Scher}, we can transform from the  $(\kappa_0, \kappa_1)$ parameters to $w$ and $w'\equiv dw/d\ln a$ evaluated at the present:
\bea \label{translation}
w(z = 0) & = &\ -1 + \frac{2\, \OO \kappa_0^2}{3}, \\
\left. \frac{d w}{d \ln a}\right|_{z=0} & = & \ \frac{2}{3} \kappa_0^2 \OO[3 - \OO (3+4\kappa_1)] \, . \nonumber
\eea
A linear model with the same values of $(w,w')$ at $z=0$ follows our model at low redshifts, but tends to diverge significantly by $z=1$, as can be seen in Fig. 1.  Note, matching at $z=0$ makes the differences between the models more apparent than if they were matched at a higher redshift; the model differences are generally small, particularly when translated into integrated observables like the luminosity distance.

\begin{figure}[htbp]
\begin{center}
\includegraphics*[width=2.5in]{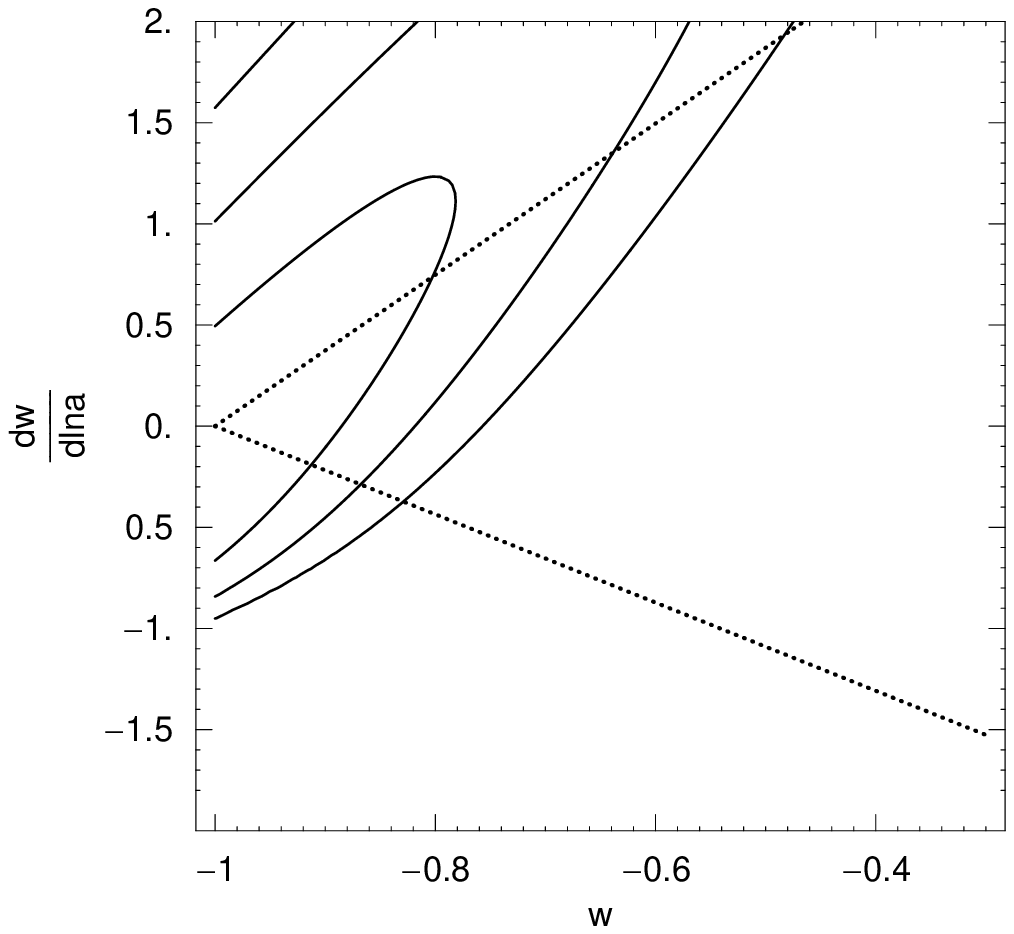}%
\hfil
\includegraphics*[width=2.5in]{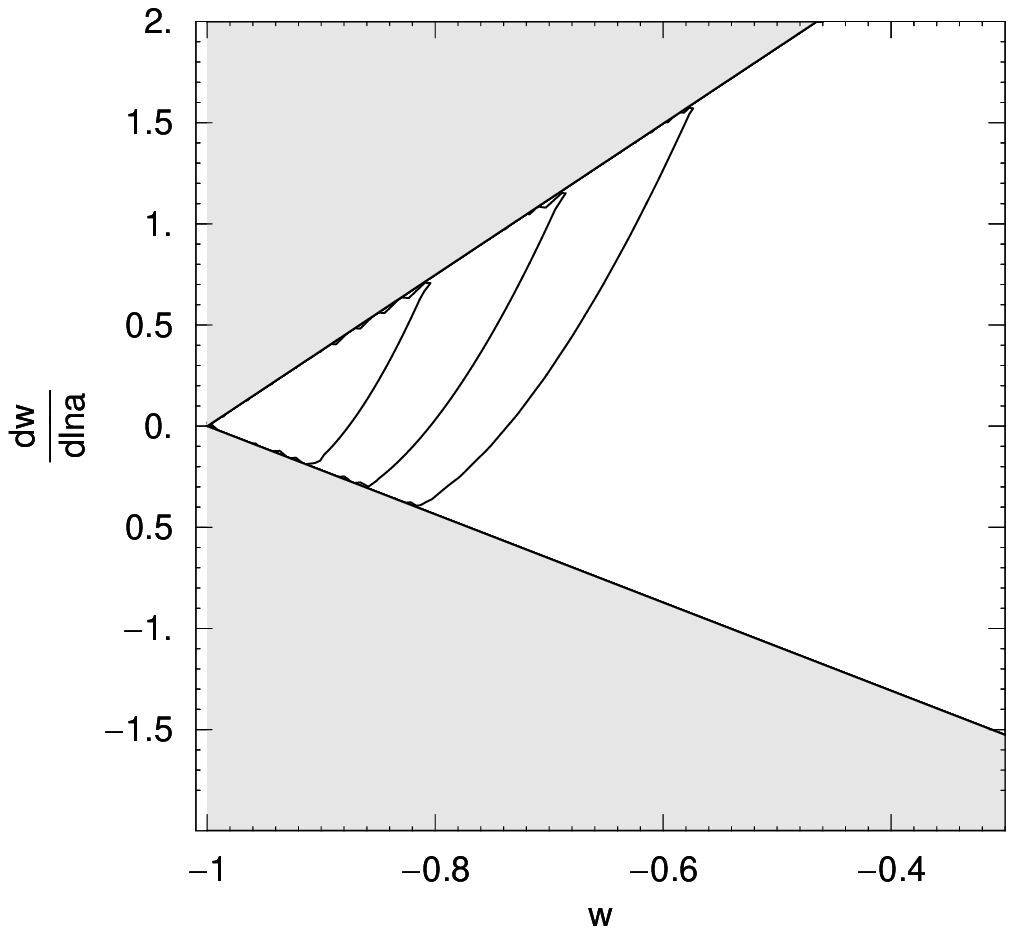}%
\end{center}\vspace{-0.3cm}
\caption{We show contours of constant likelihood resulting from SN, CMB and baryon oscillation data. 
The upper figure assumes a linear model for the equation of state of the dark energy, while the bottom figure shows our scalar field motivated parameterization, where the space of models is reduced.  The models have the same equation of state and 
derivative at $z=0$, but evolve differently at higher redshift.  For simplicity we have fixed $\OO = 0.74$.}\vspace{-.4cm}
\label{fig:likelihood}
\end{figure}

In Fig. \ref{fig:likelihood} we compare directly the observational limits for our parameterization to those of the linear parameterization.  We focus on data constraining the background evolution of the dark energy: the supernovae type Ia (SNe Ia) set of the SNLS survey \cite{Astier:2005qq} as standard candles, the baryon oscillation (BAO) peak detected in the SDSS luminous red galaxies (LRGs) \cite{Eisenstein:2005su} and the CMB shift parameter from the WMAP 3-year data \cite{Spergel} calculated by Wang and Mukherjee \cite{Wang:2006ts} as standard rulers.  These directly probe the cosmological expansion through the luminosity distance at low redshifts and the angular diameter distance at higher redshifts.  The likelihood contours for the two parametrizations are very similar with the present data, reflecting their similarity in the low redshift region where the observations are most sensitive.  
  
The likelihood curves in Fig. \ref{fig:likelihood} ignore any prior information about the models, effectively show the curves which would result from a flat prior in the $(w,w')$ space.  However, one might argue that it is more realistic to take priors in the space of potentials, e.g. assuming a uniform distribution in the derivatives of the potential or perhaps the log of the potential.  In the limit of small $\beta$, the latter corresponds to 
a uniform prior on ($\kappa_0,\kappa_1$).  Thus it is interesting to investigate how this kind of 
prior would translate into the $(w,w')$ space.  

This prior is straightforwardly given by the Jacobian of the transformation \eqref{translation}, which can be easily shown to be proportional to $[\OO(1+w)]^{-3/2}$.   Thus, given a flat field space prior, the probability diverges as one approaches $w=-1$.  While one might initially see such a prior as problematical, in fact this is not the case.  The condition $V''(\phi) \lesssim H_0^2$ effectively puts bounds on $\kappa_1$ and restricts the interesting range of the $(w,w')$ plane.  Constant $\kappa_1$ translates into straight lines converging at the cosmological constant point, as shown in Figs. 2 and 3.  Integrating the prior between such lines always results in finite answers.  Also, while the prior will tend to pull the peak of the posterior towards $w=-1$, if the data clearly prefer a higher value, the divergence in the prior will be suppressed by the exponential falloff of the likelihood.    

\begin{figure}[htbp]
\begin{center}
\includegraphics*[width=2.5in]{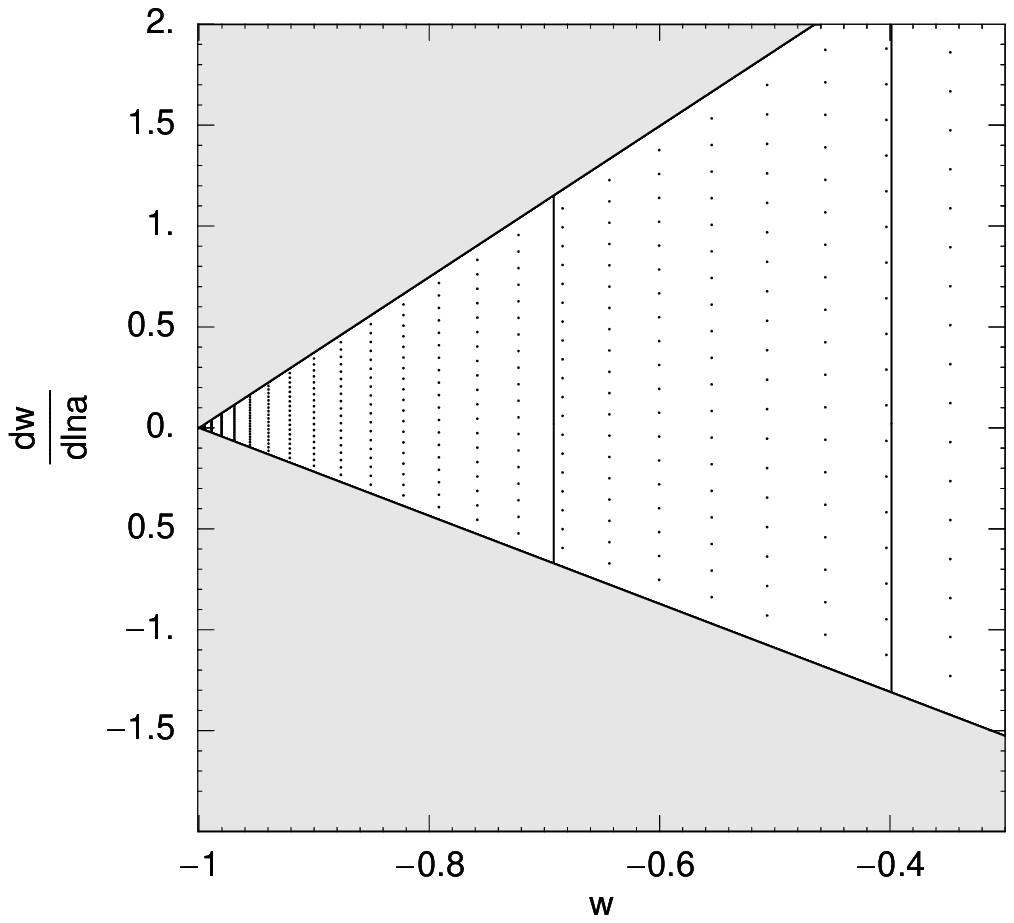}%
\hfil
\includegraphics*[width=2.5in]{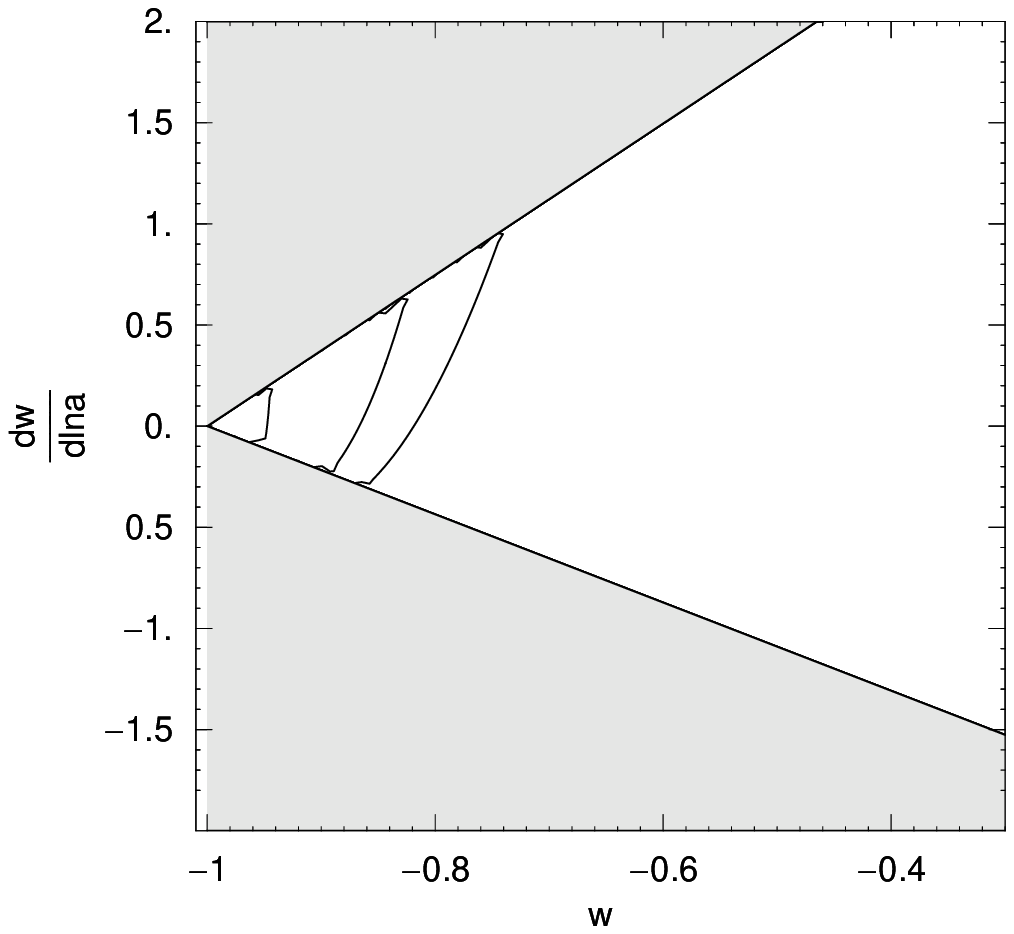}%
\end{center}\vspace{-0.3cm}
\caption{The upper figure shows the prior on $(w,w')$ space arising from a uniform prior in $(\kappa_0,\kappa_1)$ space, which diverges as $(1+w)^{-3/2}$.  The points show how a uniform grid in $(\kappa_0,\kappa_1)$ is projected into the $(w,w')$ space, while the lines show 1 and 2$\sigma$ confidence limits, assuming $w < -1/3$.   The bottom panel shows the final 
posterior, where the prior is folded in with the likelihood from the data (from Fig. 2b.)  
}\vspace{-.4cm}
\label{fig:posterior}
\end{figure}

Fig. \ref{fig:posterior} shows the prior and posterior for this parameterization, assuming a flat prior on the $(\kappa_0,\kappa_1)$ plane. 
In order to get confidence contours for the prior, we cut it off at $w=-1/3$, which corresponds to the border of the accelerating models and is also roughly where $\kappa_0 = 1$.  Both the prior and the likelihood peak around $w=-1$, making the posterior even more peaked. 
Our primary aim is to show how to translate priors from variables related to the theory to more observational parameters.  The true nature of the prior on the $(\kappa_0,\kappa_1)$ plane is still an open question and requires a better defined theory; the posterior shown in Fig. 3 should be only taken as representative of the kind of focussing that might occur.       

We have shown that as one approaches cosmological constant behaviour, the evolution of quintessence dark energy is more and more constrained by the requirement of a smooth potential, and the number of free parameters shrinks.  This provides a unique template of models which can be used to search for small deviations from the cosmological constant behaviour. 
Our parameterization is exact in this limit, and so is more appropriate than a linear model which can behave unphysically at high redshift; other parameterizations, such as a subset of the `kink' models \cite{kink}, would be an improvement to the linear models, but they are still not exact and are harder to relate to the underlying theory.    
Our approach is similar in spirit to the flow equation approach of Huterer and Peiris \cite{Huterer}, which appears to find thawing models less likely.  The reason for this discrepency is not obvious, but it likely lies in the initial conditions or in a different assumptions about the possible potentials.  

Since it is well known that the assumptions about $w(z)$ in an analysis can dramatically affect one's 
conclusions about dark energy \cite{Bassett}, 
it would be very interesting to repeat projections for future experiments using this more physical parameterization.  It seems likely that, given the convergence towards $w=-1$ at high redshift for our parameterization, experiments at low and intermediate redshift will increase in importance.  It would also be interesting to extend our approach, such as to models where the dark energy is coupled to dark matter \cite{Amendola}.

We thank Luca Amendola, Robert Caldwell, Tommaso Giannantonio, Dragan Huterer, Kazuya Koyama and 
Levon Pogosian for useful conversations. 
Research at Perimeter Institute for Theoretical Physics is supported in
part by the Government of Canada through NSERC and by the Province of
Ontario through MRI.

\end{document}